\documentclass{article}
\usepackage{spconf,amsmath,graphicx,hyperref}

\usepackage{cite}
\usepackage{amssymb,amsfonts}
\usepackage{algorithmic}
\usepackage{textcomp}
\usepackage{xcolor}
\usepackage{bm}
\usepackage{mathtools}
\usepackage{color}
\usepackage{multirow, tabu}
\usepackage{caption, hyperref, enumitem, setspace}
\usepackage{lipsum}
\usepackage{booktabs}
\usepackage[table]{xcolor}  
\usepackage{subcaption}
\usepackage{soul}

\title{Adaptive Deterministic Flow Matching for Target Speaker Extraction}
\name{Tsun-An Hsieh and Minje Kim\thanks{This material is based upon work supported by the National Science
Foundation under Grant No. 2512987.}}
\address{
 University of Illinois Urbana-Champaign, Siebel School of Computing and Data Science, USA 61801
}
\begin{document}
\ninept
\maketitle
\begin{abstract}

Generative target speaker extraction (TSE) methods often produce more natural outputs than predictive models. Recent work based on diffusion or flow matching (FM) typically relies on a small, fixed number of reverse steps with a fixed step size. We introduce Adaptive Discriminative Flow Matching TSE (AD-FlowTSE), which extracts the target speech using an adaptive step size. We formulate TSE within the FM paradigm but, unlike prior FM-based speech enhancement and TSE approaches that transport between the mixture (or a normal prior) and the clean-speech distribution, we define the flow between the background and the source, governed by the mixing ratio (MR) of the source and background that creates the mixture. This design enables MR-aware initialization, that the model starts at an adaptive point along the background–source trajectory rather than applying the same reverse schedule across all noise levels. Experiments show that AD-FlowTSE achieves strong TSE with as few as a single step, and that incorporating auxiliary MR estimation further improves target speech accuracy. Together, these results highlight that aligning the transport path with the mixture composition and adapting the step size to noise conditions yields efficient and accurate TSE. Sound examples and source codes can be found at \href{https://minjekim.com/research-projects/AD-FlowTSE#icassp2026}{https://minjekim.com/research-projects/AD-FlowTSE\#icassp2026}.

\end{abstract}
\section{Introduction}
\label{sec:intro}
In many real-world audio scenarios, the ability to extract a specific voice from a mixture is crucial. Target Speaker Extraction (TSE) addresses this need by isolating the speech of a known speaker from overlapping speech and noise using a short enrollment segment as reference. Applications include improving speech quality and intelligibility in communication systems, enhancing hearing aids, and boosting the robustness of downstream tasks such as automatic speech recognition and speaker identification.

Traditional approaches to TSE have largely been discriminative/predictive models, which directly map a noisy input to a clean output or a mask by optimizing signal-level metrics. Examples include \cite{delcroix_tdSpkBeam} and \cite{ge2020spex+}, both of which condition on an enrollment waveform to jointly enhance and identify the target speaker. While effective, these methods often introduce artifacts and can struggle to generalize to unseen speakers or noise conditions.

Recently, generative models have emerged as a powerful alternative, particularly diffusion- \cite{ho2020denoising} and flow matching (FM)-based \cite{lipman2023flow} approaches that reconstruct clean signals via learned reverse processes. In speech enhancement (SE), it is first proposed to integrate diffusion with SE by defining the forward process between clean speech and a combination of Gaussian and background noise \cite{lu2022cdiffuse}; similarly, a different diffusion process between noisy and clean speech is introduced in \cite{richter2023speech}, and this line further extends to speech restoration with architectural changes \cite{lemercier2023analysing, richter2024causal}. While effective, diffusion models are often sample-inefficient, typically requiring around 50 number of function evaluations (NFEs) to attain high quality. By contrast, recent studies indicate that FM and adaptive reverse sampling can reduce NFEs via simpler trajectories. For example, in \cite{flowse2025lee}, the FM-based SE is comparable to its diffusion counterpart with only five NFEs, and by associating the diffusion time step with signal-to-noise ratio (SNR), an SNR-aligned conditioning \cite{snr25jun} is able to achieve an efficient one-step inference with SNR prediction.
\begin{figure}[t]
    \centering
    \includegraphics[width=0.9\linewidth]{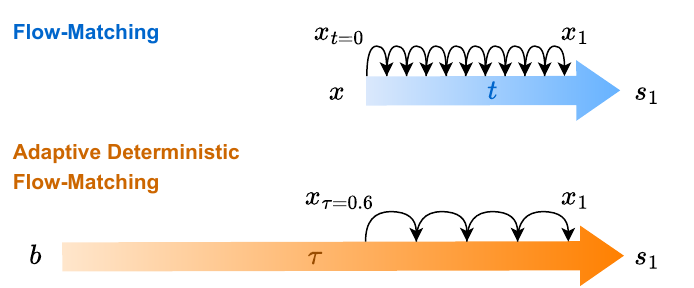}
    \caption{Comparison of FM-based TSE and our AD-FlowTSE frameworks. $x$, $s_1$, and $b$ are the mixture, target speech, and background, respectively. Depending on the quality of the input $x$, the proposed AD-FlowTSE estimates the mixing ratio $\tau$ and applies correspondingly fewer steps. Conversely, the flow matching method applies the pre-defined number of inferences inefficiently.}
    \label{fig:overview}
\end{figure}
In TSE, diffusion and FM-based methods also show promising potential. A conditional diffusion framework for TSE is proposed in \cite{kamo23_interspeech} with conditions on the mixture and an enrollment utterance, using ensemble inference to reduce extraction errors.
Discriminative and generative training objectives can be combined to improve accuracy and perceptual quality \cite{zhang2024ddtse}. Large pretrained diffusion models have been shown to be powerful in handling multiple tasks \cite{ku2025generative, liu2024generative}, including TSE and speech separation, directly operating in the frequency domain without relying on vocoders. It is shown that one NFE with an FM-based TSE is promising on the Mel-spectrogram domain, followed by a vocoder to transform the outputs into the time domain \cite{navon2025flowtse}.

Despite these advances, many approaches rely on complex pipelines or pretrained components, limiting scalability and deployment.
Instead of adding Gaussian noise to the data, cold diffusion \cite{cold2023bansal} relaxes this theoretical limitation by using multiple degradation functions in the forward process, demonstrating that diffusion models are able to learn various mapping functions between the degradation and data distributions.
Then, cold diffusion and unfolded training are introduced to SE to achieve comparable performance with discriminative models with one-step reverse process \cite{yen2023colddiffusionspeechenhancement}. Furthermore, redesigning the cold diffusion concept supports scalable and efficient inference for low-resource devices by adopting milestone tasks to simplify the reverse process and introducing a residual learning framework \cite{KimM2024sese}.

In this paper, we introduce Adaptive Deterministic flow matching for TSE (AD-FlowTSE). In line with generalizations such as \cite{yen2023colddiffusionspeechenhancement, KimM2024sese} that are based on deterministic noise input and a fixed number of function evaluations (NFE) or a fixed step size (in one-step regimes) across input qualities, AD-FlowTSE reformulates TSE within the FM paradigm. Specifically, unlike prior studies that learn transport between noisy and clean speech distributions, AD-FlowTSE learns the transport between the background (i.e., interference and/or noise) and the source (i.e., target speech) by replacing the diffusion/FM time variable with the mixing ratio (MR) between the background and source signals. In our setup, AD-FlowTSE requires test-time MR estimation to determine where a test sample lies along the learned trajectory. As shown in Fig \ref{fig:overview}, by adapting the FM paradigm, our model affords dynamic control over the number of reverse steps or the step size, enabling inference-time flexibility. Notably, with as few as a single step, our results show that AD-FlowTSE achieves the state-of-the-art generative TSE performance on Libri2Mix \cite{CosentinoJ2020librimix}, rivals larger pretrained FM models, and improves the accuracy of extracting the intended target speaker.

\section{Background}
\subsection{Target Speaker Extraction}
TSE aims to recover the target speech indicated by an enrollment utterance. Let the mixture be
\begin{equation}
x = \tau s_1 + (1-\tau)b, \quad \tau\in[0,1]
\label{eq:prob_def}
\end{equation}
where $\tau$ is the MR; $s_1$ is the clean target speech; $\{s_i\}_{i=2}^{N}$ are $N-1$ interfering speakers; $n$ is a non-speech interference, and, therefore, $b=\alpha_1n + \sum_{i=2}^N\alpha_is_i$ denotes the background weighted by $\{\alpha_i\}_{i=1}^{N}$. Given the mixture $x$ and an enrollment $e$ (in waveform or time–frequency domain) that specifies the target speaker $s_1$, we learn a TSE mapping $f$ such that
\begin{equation}
s_1\approx \hat{s}_1 = f(x,e).
\end{equation}
This formulation makes explicit that TSE extracts only the target component $s_1$ from $x$ while treating the sum of all remaining speech and non-speech interferences (including noise) as the background source $b$.

\subsection{flow matching for TSE}
Conditional FM (CFM) for TSE specifies a time-indexed conditional probability path $p_t(x_t | x, s_1)$, $t\in[0,1]$, that interpolates between the mixture distribution at $t=0$ and the target speech distribution at $t=1$, conditioned on the observed mixture $x$ and $s_1$. 
The path is realized by a probability-flow ordinary differential equation (ODE) whose flow map $\psi_t$ transports the mixture distribution $p_0$ to $p_t$. From this, we derive the time-dependent target velocity field $u_t$ associated with the chosen conditional path:
\begin{equation}
\frac{d}{dt}\psi_t(x_t | x, s_1) = u_t(x_t|x, s_1),
\quad x_t \sim p_t, \quad x_1\approx s_1.
\end{equation}
Training regresses a neural field $v_\theta$ parameterized by $\theta$ toward this target at samples $x_t\sim p_t(x_t | x, s_1)$:
\begin{equation}
\mathcal{L}_{\mathrm{CFM}}
=
\mathbb{E}_{(x,e, s_1)}
\int_0^1
\mathbb{E}_{x_t\sim p_t}
\Vert
v_\theta (x_t, e, t) - u_t(x_t | x, s_1)
\Vert_2^2 dt,
\end{equation}
and sampling integrates $dx_t = v_\theta(x_t, e, t)dt$ from $x_0\sim p_0$ to obtain a draw at $t=1$.

Given supervised TSE triples $\{(x,e,s_1)\}$ (mixture $x$, enrollment $e$, clean target $s_1$), one can instantiate $p_t(x_t | x, s_i)$ with a Gaussian conditional path whose mean follows a mixture-to-target interpolation with a scale that shrinks over time:
\begin{equation}
\mu_t(x, s_1) = (1-t)x + ts_1,
\quad
\sigma_t = (1-t)\sigma_{\mathrm{max}} + t\sigma_{\mathrm{min}},
\end{equation}
so that
\begin{equation}
p_t(x_t | x, s_1) = \mathcal{N}\big(x_t; \mu_t, \sigma_t^2 I \big).
\end{equation}
For this Gaussian family, the associated probability-flow velocity has the closed form
\begin{equation}
u_t(x_t|x,s_1) = 
\frac{\sigma_t'}{\sigma_t}(x_t - \mu_t(x, s_1)) + \mu_t'(x, s_1),
\quad
(\cdot)'=\frac{d(\cdot)}{dt}
\label{eq:gaussian_path}
\end{equation}
which specializes to a deterministic endpoint at $t=1$ if $\sigma_{\mathrm{min}}=0$.

The estimator is conditioned on the observation and enrollment through its inputs, $v_\theta(x_t, e, t)$, and train with the optimal transport CFM (OT-CFM) objective \cite{tong2024improving}
\begin{equation}
    \mathcal{L}_{\mathrm{OT\text{-}CFM}} = 
    \mathbb{E}_{(x, e, s_1),t, x_t}
    \Vert
    v_\theta(x_t, e, t) - u_t(x_t|x, s_1)
    \Vert_2^2,
\end{equation}
At inference, we draw $x_0 \sim p_0$ and integrate the ODE from $t=0$ to $t=1$ to produce a sample of the target speech conditioned on $e$.

\subsection{Limitation}
\label{ssec:limit}
Under a general FM framework, an ODE sampler employs a fixed unit-time transport $t\in[0,1]$ for every input, regardless of the mixture’s quality, i.e., the MR between the target source and the background. This design reflects that FM was initially developed as a generative model to synthesize signals (i.e., images, audio, speech, etc.) from an easy-to-sample distribution, e.g., the standard normal.

In the context of TSE, however, each mixture $x$ effectively lies at a point along the conditional flow $(1-\tau)b + \tau s_1$ defined between the target speech $s_1$ and the background $b$, where $\tau$ serves as the MR that generates $x$. Forcing the model to traverse the same full interval $[0,1]$ can therefore be suboptimal. For example, when the inputs are cleaner (large $\tau$), integrating the entire span wastes computation and may hallucinate; when the inputs are noisier (small $\tau$), a fixed-step budget may be insufficient to reach the target endpoint. This mismatch increases optimization difficulty and sampling inefficiency.

To address this issue, in Sec.~\ref{sec:method} we present an adaptive FM scheme that flows between $b$ and $s_1$. Our model not only learns $v_\theta$, but also estimates $\hat{\tau}\approx\tau$ from $(x,e)$ at inference, and initializes at $x_{\hat{\tau}}=x$, and integrates only over the residual interval $[\hat \tau,1]$ to allocate an adequate number of steps or a step size proportional to $1-\hat \tau$, which associates the trajectory length to the mixture’s MR.




\section{Adaptive Deterministic FM for TSE}
\label{sec:method}

\subsection{MR-Informed Vector Field Estimator Training}
To address the issues described in Sec.~\ref{ssec:limit}, we thereby present AD-FlowTSE that flows from the background distribution to the target speech distribution as defined in Eq. \ref{eq:prob_def}, where we ties the trajectory directly to mixture composition.
Then, we instantiate the conditional probability path by
\begin{equation}
\mu_\tau(b,s_1) = (1-\tau)b + \tau s_1, \quad
\sigma_\tau = (1-\tau)\sigma_{\max} + \tau\sigma_{\min},
\label{eq:mr-path}
\end{equation}
and define $p_\tau(x_\tau |b,s_1) = \mathcal{N}\big(x_\tau;\mu_\tau(b,s_1),\sigma_\tau^2 I\big)$.

We follow the cold-diffusion/SESE formulation \cite{yen2023colddiffusionspeechenhancement, KimM2024sese}, in which the denoising trajectory between noisy and clean speech is defined by a sequence of predefined, task-specific degradations or milestones rather than by Gaussian noise injection.
Therefore, we adopt a deterministic reverse process by specifying both $\sigma_{\min}$ and $\sigma_{\max}$ to 0.
This choice is based on the assumption that the mixture is defined as a linear combination of $s_1$ and $b$ if no nonlinear acoustic processes were involved, which is also observed in \cite{modifying2025korostik} where more steps are required for dereverberation tasks. 
Hence $\tau=0$ corresponds to the background endpoint, while $\tau=1$ coincides with the target endpoint, i.e., $x_1 = s_1$ with no stochasticity, i.e., $\sigma_{\min}=0$.
Plugging $\mu_{\tau}$ and $\sigma_{\tau}$ into Eq. \ref{eq:gaussian_path}, the associated target vector field becomes 
\begin{equation}
    u_\tau(x_\tau|b, s_1) = s_1 - b.
\end{equation}

We train a neural velocity field $v_\theta(x_\tau, e,\tau)$ to match $u_\tau$ by minimizing the OT-CFM loss:
\begin{equation}
\mathcal{L}_{\mathrm{OT\text{-}CFM}}(\theta)
=
\mathbb{E}_{(b, s_1,e),\tau,x_\tau}
\Vert
v_\theta(x_\tau, e, \tau) - u_\tau(x_\tau|b,s_1)
\Vert_2^2,
\label{eq:mr-cfm}
\end{equation}
where, during supervised training, $(b, s_1, 3)$ are drawn from the data-generation process and satisfy \eqref{eq:prob_def} with an arbitrary $\tau\sim\mathcal{U}[0,1]$. Equation \eqref{eq:mr-cfm} learns a condition- and MR-aware velocity that transports mass from background to target.

\subsection{MR Predictor Training}
During training of the velocity estimator $v_\theta$, the ground-truth MR $\tau$ is known from the data-generation process and is used to construct the path states (i.e., to remix background and target along the designed trajectory), while the MR is unknown at inference. 
To resolve this, we learn a standalone MR regressor $g_\phi(x,e)$ parameterized by $\phi$ that predicts ${\tau}\approx\hat{\tau}\in[0,1]$ from the mixture $x$ and enrollment $e$. 

Concretely, we adopt ECAPA-TDNN \cite{desplanques2020ecapa} as the backbone, and apply a shared feature extractor $w(\cdot)$ to $x$ and $e$. Next, we concatenate the embeddings, and map them to a scalar via a small multi-layer perceptron $h(\cdot)$ with a sigmoid activation function $\sigma(\cdot)$ to enforce the $[0,1]$ output range:
\begin{equation}
\hat{\tau} = g_\phi(y, e) = \sigma(h([w(y);w(e)])),
\end{equation}
where $[\cdot; \cdot]$ denotes concatenation. 
The MR head is trained with mean-squared error against the ground-truth MR $\tau$ by minimizing $\mathcal{L}_{\mathrm{MR}}(\phi)=\mathbb{E}\big[(\hat{\tau}-\tau)^2\big]$.

\subsection{AD-FlowTSE Inference}
At test time, $\hat{\tau}$ seeds the adaptive integration schedule:
\begin{equation}
    x_{\hat\tau_{j+1}} = 
    \Delta \hat\tau_j \cdot v_\theta(x_{\hat\tau_j}, e, \hat\tau_j)+ x_{\tau_j},
    \quad
    \hat\tau_0=\hat\tau,
\end{equation}
where $\Delta \hat\tau_j=\hat\tau_{j+1} - \hat\tau_j$, and $j$ denotes the step index. 

By iterating the reverse steps proportional to $1-\hat{\tau}$, the mixture $x_{\hat{\tau}}=x$ ideally converges to the clean target $s_1=x_1$. This adaptive schedule shortens the trajectory for high-SNR inputs (large $\hat{\tau}$) and lengthens it for low-SNR inputs (small $\hat{\tau}$), aligning computational complexity with mixture quality.

\section{Experiments}
\subsection{LibriMix Dataset}
For TSE, we follow the setup in \cite{Delcroix2018speaker_beam}, which uses the Libri2Mix dataset, a derivative of LibriSpeech designed for multi-speaker separation and extraction.
The Libri2Mix dataset was used for training and evaluation. Training data included both the \texttt{train-360} and \texttt{train-100} subsets, with \texttt{dev} used for validation and \texttt{test} for final evaluation. We employed 6-second input segments comprising 3-second enrollment and 3-second mixture, sampled at 16kHz. The STFT is applied with a window length and $n_\text{fft}$ of 510, alongside a hop length of 128.

\subsection{Training Configuration}
We train the proposed AD-FlowTSE model using a batch size of 64, and run for up to 2000 epochs. For inference, we use the best checkpoint with the highest validation SI-SDR. Mixed precision training is adopted (16-bit) to reduce memory usage and accelerate training. Training was performed on 16 GPUs with distributed data parallelism and gradient clipping set to 0.5 to prevent exploding gradients. The learning rate was initially set to $1\times10^{-4}$ and scheduled using the cosine annealing policy with a minimum learning rate of $1\times10^{-5}$, a warm-up period of 5 epochs, and $T_{\max} = 50$. Weight decay was set to 0.01. The AdamW \cite{loshchilov2017decoupled} optimizer was used throughout. 
We employ a UNet-style DiT (UDiT) \cite{liu2024generative} model that consists of 16 transformer \cite{VaswaniA2017nips} layers with 16 attention heads and a hidden size of 768. The input and output dimensions were both 512. No positional encoding was applied, and the positional length was set to 500. The solver used an Euler method with 1,000 maximum steps for training and 1--20 maximum steps for evaluation.

\subsection{Baseline Models}
We benchmark AD-FlowTSE primarily against the results reported in \cite{navon2025flowtse}, which provide a comprehensive evaluation on both the clean and noisy configurations of the Libri2Mix dataset. In particular, we compare our model and its variants to both discriminative and generative approaches
We consider DiffSep+SV and DDTSE \cite{zhang2024ddtse}, DiffTSE \cite{kamo23_interspeech}, and FlowTSE \cite{navon2025flowtse}, the last of which is most closely related to our method. We also include SR-SSL \cite{ku2025generative}, which follows a pretraining–finetuning paradigm, and SoloSpeech \cite{wang2025solospeech}, which employs a UDiT-based architecture aligned with our model class.

\begin{table*}[t]
\centering
\setlength{\tabcolsep}{4pt}
\resizebox{.95\textwidth}{!}{%
\begin{tabular}{lcccccccccccccc}
\toprule
\multirow{2}{*}{Method} & \multirow{2}{*}{Type} & \multicolumn{6}{c}{Libri2Mix Noisy} & & \multicolumn{6}{c}{Libri2Mix Clean} \\
\cmidrule(lr){3-8} \cmidrule(lr){10-15}
  &  & PESQ & ESTOI & SI-SDR & OVRL & DNSMOS & SIM & & PESQ & ESTOI & SI-SDR & OVRL & DNSMOS & SIM \\
\midrule
Mixture    & -- & 1.08 & 0.40 & -1.93 & 1.63 & 2.71 & 0.46 &  & 1.15 & 0.54 & 0.00 & 2.65 & 3.41 & 0.54\\
\midrule
DiffSep+SV \cite{zhang2024ddtse} & \multirow{6}{*}{G} & 1.32 & 0.60 & -- & 2.78 & 3.63 & 0.62 &  & 1.85 & 0.79 & -- & 3.14 & 3.83 & 0.83 \\
DDTSE \cite{zhang2024ddtse} &  & 1.60 & 0.71 & -- & 3.28 & 3.74 & 0.71 &  & 1.79 & 0.78 & -- & \textbf{3.30} & \textbf{3.79} & 0.73 \\
DiffTSE \cite{kamo23_interspeech} &  & -- & -- & -- & -- & -- & -- &  & 3.08 & 0.80 & 11.28 & -- & -- & -- \\
FlowTSE \cite{navon2025flowtse} &  & 1.86 & 0.75 & -- & \textbf{3.30} & \textbf{3.82} & 0.83 &  & 2.58 & 0.84 & -- & 3.27 & \textbf{3.79} & 0.90 \\
SR-SSL \cite{ku2025generative} &  & -- & -- & -- & -- & -- & -- &  & \textbf{2.99} & -- & 16.00 & -- & -- & -- \\
SoloSpeech\textsuperscript{\textdagger} \cite{wang2025solospeech} &  & 1.89 & 0.78 & 11.12 & -- & 3.76 & -- &  & -- & -- & -- & -- & -- & -- \\
\midrule
Ours: Estimated $\hat{\tau}$ & \multirow{3}{*}{D} & \textbf{2.15} & \textbf{0.81} & \textbf{12.69} & 3.11 & 3.48 & \textbf{0.87} &  & 2.89 & \textbf{0.90} & \textbf{17.49} & 3.15 & 3.59 & \textbf{0.95} \\
Ours: Oracle $\tau$       &  & 2.16 & 0.81 & 12.85 & 3.11 & 3.48 & 0.87 &  & 2.92 & 0.90 & 17.73 & 3.16 & 3.60 & 0.95 \\
Ours: Random $\tilde{\tau}$ &  & 1.93 & 0.74 & 9.14 & 2.97 & 3.37 & 0.85 &  & 2.57 & 0.83 & 13.26 & 3.09 & 3.55 & 0.93 \\
\midrule
Ours: $\tau=1$ & \multirow{2}{*}{D} & 1.08 & 0.40 & -1.93 & 1.63 & 2.71 & 0.72 &  & 1.15 & 0.54 & 0.00 & 2.65 & 3.41 & 0.76 \\
Ours: $\tau=0$ &  & 1.73 & 0.72 & 9.40 & 2.87 & 3.23 & 0.84 &  & 2.33 & 0.82 & 12.54 & 3.02 & 3.44 & 0.92 \\
\bottomrule
\end{tabular}
}
\caption{Performance comparison on the Libri2Mix Noisy and Clean sets. \textsuperscript{\textdagger}For consistency, we omit the SIM scores reported in \cite{wang2025solospeech} because they use a different speaker verification backend.}
\label{tab:performance_comparison}
\end{table*}

\subsection{Evaluation Metrics}
We evaluate both perceptual quality and target identity using five popular metrics: perceptual evaluation of speech quality (PESQ) \cite{RixA2001pesq}, extended short-time objective intelligibility (ESTOI) \cite{TaalC2010icassp}, scale-invariant signal-to-distortion ratio (SI-SDR) \cite{LeRouxJL2018sisdr}, OVRL, DNSMOS \cite{reddy2021dnsmos} and the speaker similarity (SIM) \cite{wang2023wespeaker} to quantify speaker identity preservation, computed as the cosine similarity between embeddings of the estimated and target signals.

\section{Results}
\subsection{Overall Performance}
Table \ref{tab:performance_comparison} compares our method against other TSE systems on the noisy and clean subsets of Libri2Mix. Overall, our approach consistently outperforms the competing generative baselines in terms of the intrusive metrics (i.e., PESQ, ESTOI, and SI-SDR), indicating better perceptual quality, intelligibility, and signal fidelity. This gain comes with a modest trade-off on the non-intrusive measures (i.e., OVRL and DNSMOS), where some methods remain slightly higher. Crucially, our model achieves noticeably stronger SIM, reflecting more reliable target speaker preservation. Compared to SR-SSL with 430M parameters, our model (including $v_\theta$ and $g_\phi$) achieves better performance while using only 83\% as many parameters. Overall, these results show that adaptively seeding the mixture with $\hat{\tau}$ yields the state-of-the-art performance in intrusive measures, and markedly improved extraction accuracy, with only minor concessions in non-intrusive scores.
\begin{figure}[t]
    \centering
    \includegraphics[width=\linewidth]{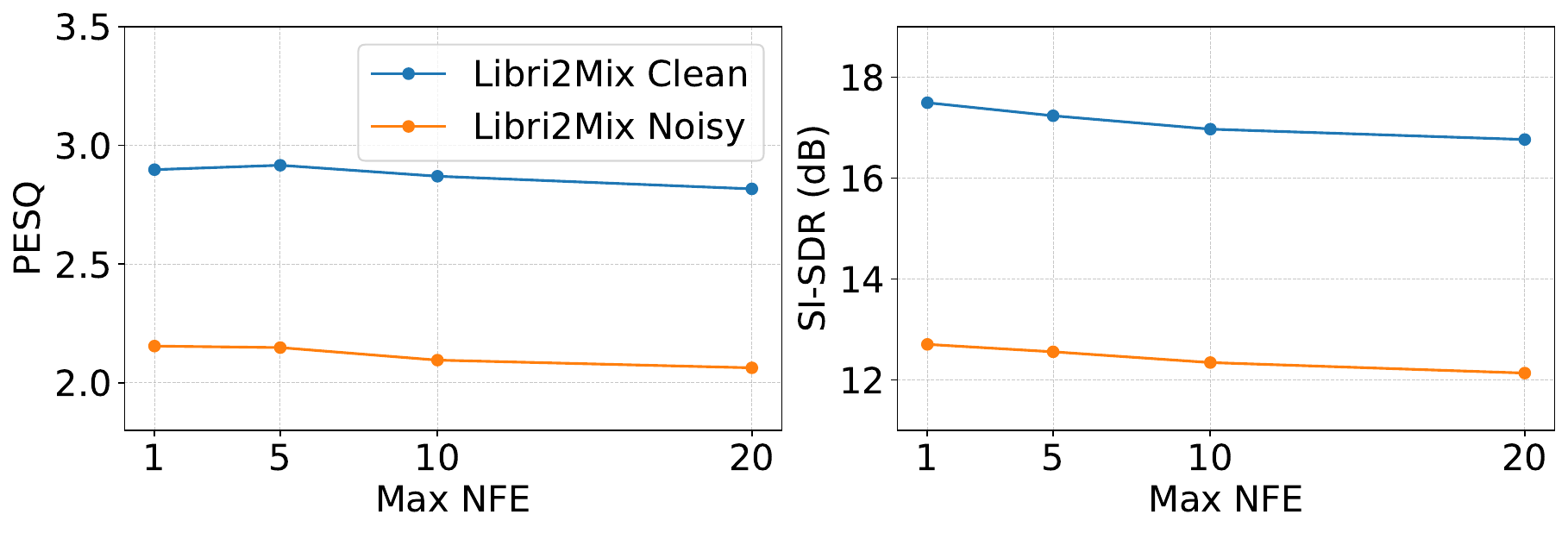}
    \caption{Performance across NFE.}
    \label{fig:performance_by_nfe}
\end{figure}
\subsection{Robustness to Inaccurate MR Estimation}
Next, we compare three sampling schedules based on the MR $\tau$. The oracle setup uses the ground truth $\tau$ as an upper bound on performance for a fixed architecture and solver, representing the best we can do when the flow time perfectly matches the mixture composition. 
Our estimated setup replaces $\tau$ with $\hat{\tau}$ inferred from $g_\phi(x,e)$ and tracks the oracle closely across all metrics, indicating that the MR estimator is accurate enough to leverage most of the benefit of adaptive FM. In contrast, using a random $\tilde{\tau} \sim \mathcal{U}[0,1]$ substantially degrades results, showing that a certain level of accuracy of $\tau$ estimation is needed for both the effectiveness of signal extraction and target preservation.
Finally, we probe failure modes by fixing $\tau=1$ and $\tau=0$. Setting $\tau=1$ assumes the mixture already lies at the target endpoint, so the sampler performs no correction. Consequently, all scores remain unchanged from the input, except for SIM.
Conversely, $\tau=0$ forces the model to traverse the entire background-to-target path regardless of input cleanliness, systematically overshooting and introducing artifacts and hallucinations. These extremes indicate that accurate $\tau$ estimation is critical, as miscalibration either attenuates the benefit of adaptation ($\tau=1$) or over-amplifies it ($\tau=0$), degrading overall performance.

\subsection{NFE Analysis}

Fig. \ref{fig:performance_by_nfe} shows how performance changes as we increase the maximum NFE (Max NFE) in the esitmate MR setup. For both subsets, the best scores occur at NFE=1 or 5 with minor differences and then gradually decline as NFE grows. This indicates that, with our MR-aware initialization, the remaining transport from the mixture to the target only requires a few steps, and forcing longer trajectories adds accumulation error and tends to over-correct, yielding worse PESQ and SI-SDR. A single step already captures the needed update, so additional steps provide no benefit and mildly hurt quality, highlighting both the efficiency and the accuracy of the adaptive one-step regime.

\section{Conclusion}
We presented AD-FlowTSE, a FM formulation of TSE that aligns the transport path length with the MR. Instead of defining the flow between the mixture and the target speech, AD-FlowTSE learns a deterministic flow from the background to the target by reinterpreting the flow time as the MR. At inference, an MR predictor enables MR-aware initialization at $x_{\hat{\tau}}=x$ and adaptive integration only over $[\hat{\tau},1]$ with a step budget proportional to $1-\hat{\tau}$.
Experimental results indicates that AD-FlowTSE consistently outperforms prior systems on intrusive metrics while achieving a notably higher SIM, with only modest trade-offs on the non-intrusive scores. Our NFE analysis further shows that the MR predictor already captures the required amount to update, and longer trajectories provide no benefit and can even degrade quality, underscoring the efficiency of the proposed adaptive scheme. Overall, the results validate that replacing diffusion time with MR, coupled with MR-aware initialization and adaptive reverse integration, yields an accurate and computationally efficient approach to TSE.

\bibliographystyle{IEEEbib}
\bibliography{strings,refs}

\end{document}